\documentclass[twocolumn,prl,showpacs,superscriptaddress]{revtex4}

\usepackage{amsfonts}
\usepackage{amsmath}
\usepackage{amssymb}
\usepackage{graphicx}

\begin{document}

\preprint{HEP/123-qed}
\title[Nonlinear Wigner Solid Transport]{Nonlinear Wigner
Solid Transport Over Superfluid Helium Under AC Conditions}
\author{Yuriy P. Monarkha}
\affiliation{Low Temperature Physics Laboratory, RIKEN, Hirosawa
2-1, Wako 351-0198, Japan} \affiliation{Institute for Low
Temperature Physics and Engineering, 47 Lenin Avenue, 61103
Kharkov, Ukraine}
\author{Kimitoshi Kono}
\affiliation{Low Temperature Physics Laboratory, RIKEN, Hirosawa
2-1, Wako 351-0198, Japan}



\begin{abstract}
Nonlinear transport properties of the two-dimensional Wigner solid
of surface electrons on superfluid helium are studied for
alternating current conditions. For time-averaged quantities like
Fourier coefficients, the field-velocity characteristics are shown
to be qualitatively different as compared to that found in the DC
theory. For a spatially uniform current we found a general
solution for the field-velocity relationship which appears to be
strongly dependent on the current frequency. If the current
frequency is much lower than the ripplon damping parameter, the
Bragg-Cherenkov resonances which appear at high enough drift
velocities acquire a distinctive saw-tooth shape with long
right-side tails independent of small damping. For current
frequencies which are close or higher than the ripplon damping
coefficient, the interference of ripplons excited at different
time intervals results in a new oscillatory (in drift velocity)
regime of Bragg-Cherenkov scattering.

\end{abstract}

\maketitle

\section{Introduction}

Electrons trapped on the surface of liquid helium form a clean
two-dimensional electron system of which the average Coulomb
interaction energy can greatly exceed the average kinetic energy
(for a review, see~\cite{And-book,MonKon-book}). For typical
electron densities realized experimentally ($n_{s}\lesssim
10^{9}\,\mathrm{cm}^{-2}$), the Fermi-energy of surface electrons
(SEs) is much smaller than temperature. Therefore at a low enough
temperature, depending on the surface electron density ($
T_{c}\propto \sqrt{n_{s}}$), this electron system undergoes a
transition to the Wigner solid state. This was first observed by
Grimes and Adams~~\cite{GriAda-79} from the onset of resonances
induced by electron interaction with capillary-waves (ripplons)
whose wave-vector $\mathbf{q}$ is close to electron reciprocal
lattice vectors $\mathbf{g}$. Since the electron lattice
determines a specific set of frequencies $\omega _{n}=\omega
_{g_{n}}$ [here $ n=1,2,3,...$ , $\omega _{q}=\sqrt{\alpha /\rho
}q^{3/2}$ is the ripplon spectrum, $\alpha $ is the surface
tension, and $\rho $ is the mass density of liquid helium],
experimental evidence for a triangular electron lattice on a
liquid-He surface was given. Electron interaction with such
ripplons appeared to be very strong, leading to a huge
reconstruction of the WS phonon spectrum in the low frequency
range~\cite{FisHalPla-79}.

The resonances of Grimes and Adams occur when the frequency of the
input signal $\omega $ is close to $\omega _{n}$. Transport
properties of the WS of SEs are usually studied under much lower
frequencies ($\omega \ll \omega _{1}$). Nevertheless, even under
low frequency conditions the resonant interaction with ripplons of
frequencies which are close to $\omega _{n}$ appear to be possible
as a nonlinear conductivity effect~\cite{KriDjeFoz-96}. The
physics of this phenomenon can be explained as follows. The
pressure at the interface induced by the WS moving with a constant
drift velocity $\mathbf{v}$ can be represented as a series of
terms proportional to $\exp \left[ i \mathbf{g\cdot }\left(
\mathbf{r-v}t\right) \right] $ , where $\mathbf{r}$ is the
in-plane coordinate vector. As a function of $t$, it can be
considered as a series of harmonic perturbations with frequencies
$\mathbf{gv}$, and one can expect a resonance response of the
system when $\mathbf{gv}$ is close to $\omega _{n}$. It should be
noted that the corresponding velocity $\mathrm{v}_{1}=\omega
_{1}/g_{1}$ is rather low (typically about $ 1-10\,\mathrm{m/s}$ )
while the thermal velocity of electrons in the liquid state is
usually much higher (about $3\cdot 10^{3}\,\mathrm{m/s}$).
Therefore the single-electron Cherenkov emission of ripplons is a
quite usual phenomenon contributing into resistivity of SEs on
liquid helium. The important point is that for a moving WS with
$\mathbf{gv\rightarrow }\omega _{n}$, the response of the system
can be considered as a coherent Bragg-Cherenkov (B-C) scattering.
This effect was first described by the usual perturbation
treatment~\cite{DykRub-97} which leads to symmetrical peaks of the
electron collision rate with non-Lorentzian tails.

Besides direct B-C scattering effects limiting the WS
velocity~\cite{KriDjeFoz-96}, there are other interesting
nonlinear conductivity ($\sigma $) phenomena observed for SE
transport over superfluid helium. In the presence of the magnetic
field oriented normally to the surface, $\sigma _{xx}^{-1}$ as a
function of the input voltage $V$ has a remarkable N-type
anomaly~\cite{ShiKon-95}. The decreasing part of $\sigma
_{xx}^{-1}(V)$ was attributed to the B-C scattering. When studying
WS confined in the channel geometry, periodic conductance
oscillations with varying the drift velocity were
observed~\cite{GlaDotFoz-01}. These oscillations were attributed
to anisotropic spatial order with lines of electrons along the
channel edges. Complicated nonlinear conductivity of the WS was
observed for current frequencies which are close to typical
frequencies of plasmon-ripplon coupled modes~\cite{SyvNasNeo-08}.
Interesting WS velocity jumps caused by the decoupling of
electrons from the surface deformation were recently observed for
the WS in a channel~\cite{IkeAkiKon-09}.

Unfortunately, for $q\rightarrow g_{1}$ the electron-ripplon
coupling is strong which leads to a huge increase of the electron
effective mass at low frequencies due to surface
dimples~\cite{FisHalPla-79}. Under such conditions the
perturbation treatment is doubtful and coupled WS phonon-ripplon
modes are usually treated in a self-consistent
way~\cite{Nam-80,MonShi-83}. In this treatment the most of the
interaction Hamiltonian of the WS with ripplons of $q\rightarrow
g_{1}$ is included in the description of the coupled modes.
Therefore, a simple classical model of coherent B-C scattering of
capillary waves for the WS moving with a constant
velocity~\cite{Vin-99} seems to be more appropriate than a
perturbation treatment. This model was introduced in order to
analyze a complicated nonlinear magnetoconductivity observed
previously~\cite{ShiKon-95}. An extension of this model applied to
liquid $^{3}\mathrm{He}$ allows to explain the nonlinear
field-velocity characteristics of the WS under strong ripplon
damping conditions~\cite{ShiMonKon-04,MonKon-05}.

Application of the models of coherent B-C scattering to the
nonlinear WS transport on liquid $^{4}\mathrm{He}$ is difficult
for several reasons. First, experimental geometries usually imply
that the driving electric field and electron current density are
not spatially uniform. Secondly, measurements are done under AC
conditions when the electron velocity and driving electric field
are periodic functions of time with the period $2\pi
/\omega $ which is much longer than the typical ripplon oscillation period $%
2\pi /\omega _{1}$. Moreover, the damping of ripplons $\gamma
_{q}$ in superfluid $^{4}\mathrm{He}$ is anomalously
small~\cite{RocRogWil-96}, which means that B-C peaks of the
classical model are extremely narrow and some other effects not
included in the model can significantly affect its main results.

In this work we study the surface-displacement profile and
field-velocity relationship for spatially uniform alternating
motion of the WS over superfluid $^{3}\mathrm{He}$ and
$^{4}\mathrm{He}$ under small ripplon damping conditions: $\gamma
_{g_{1}}\ll \omega _{1}$. For arbitrary frequency of the current
$\omega$, the exact expression for the field-velocity relationship
can be found. This solution appears to be strongly dependent on
the ratio $\omega /\gamma _{g}$. Therefore, we separate two
frequency regions: $\omega \ll \gamma _{g}$ and $\omega \gtrsim
\gamma _{g}$. In both regions, the nonlinear field-velocity
characteristics obtained here for time averaged quantities differ
significantly from those obtained in the DC model. In the high
frequency region $\gamma _{g}\lesssim \omega \ll \omega _{g}$, we
expect the appearance of a new B-C scattering regime of the WS
transport caused by interference of ripplons excited at different
time intervals. This frequency region is usually realized in
experiments on nonlinear WS transport over superfluid
$^{4}\mathrm{He}$, and, therefore, we expect that our new results
will help to understand the nonlinear electronic response
observed.

\section{Dimple profile evolution induced by the drift velocity}

The analysis of the classical B-C scattering given in
Ref.~\cite{Vin-99} was restricted to a simplified one-dimensional
DC model. The damping effects were considered in a
phenomenological way. Here we consider more realistic 2D model of
alternating motion of the WS with a particular ripplon damping
defined for both $^{3}\mathrm{He}$ and $^{4}\mathrm{He}$. We
investigate shape variations induced by the WS velocity and
ripplon damping which are very important for understanding the
nonlinear WS transport.

We assume spatially uniform motion of the WS, which means that all
electron lattice sites have the same displacement vector
$\mathbf{s}(t)$ in external fields. In this case the electron
pressure at the interface induced by the WS moving with an
arbitrary velocity can be presented in the following form:
\begin{equation}
P^{(\mathrm{el})}(\mathbf{r},t)=n_{s}\sum_{\mathbf{g}}\tilde{V}_{g}\exp
\left[ i\mathbf{g}\cdot \left( \mathbf{r-s}(t)\right) \right] ,
\label{e1}
\end{equation}
where $\tilde{V}_{q}=V_{q}\exp \left( -q^{2}\left\langle
u_{\mathrm{f} }^{2}\right\rangle /4\right) $, the electron-ripplon
coupling $V_{q}$ depends on the holding electric field $E_{\bot }$
directed normally to the surface and on the wavenumber
$q$~\cite{MonKon-book}, and $\left\langle u_{
\mathrm{f}}^{2}\right\rangle $ - is the mean-square displacement
of electrons from lattice sites due to fast coupled phonon-ripplon
modes whose frequencies are limited by $\omega _{\mathrm{f}}\gg
\omega _{1}$. Actually, $\omega _{\mathrm{f}}$ is the frequency of
electron oscillations in the potential of a steady dimple. A
simple self-consistent treatment~\cite{MonShi-83} gives
\begin{equation}
\left\langle u_{\mathrm{f}}^{2}\right\rangle \simeq
\frac{\left\langle u_{0}^{2}\right\rangle +u_{T}^{2}\ln \left(
T/\hbar \omega _{V}\right) }{ 1-g_{1}^{2}u_{T}^{2}/4},  \label{e2}
\end{equation}
where $\left\langle u_{0}^{2}\right\rangle \simeq
1.248\hbar/(2m_{e}c_{t}\sqrt{\pi n_{s}})$ is the mean-square
displacement due to zero point vibrations,
\begin{equation}
u_{T}^{2}=\frac{T}{2\pi m_{e}n_{s}c_{\mathrm{t}}^{2}},\mathtt{\ \
\ }\omega _{V}=\sqrt{\frac{3n_{s}}{\alpha m_{e}}}V_{g_{1}},
\label{e3}
\end{equation}
$m_{e}$ is the free electron mass, and $c_{\mathrm{t}}^{2}$ is the
transverse sound velocity of the electron solid.

Considering ripplons as an ensemble of surface oscillators with a
certain damping parameter $\gamma _{q}$ surface displacements $\xi
(\mathbf{r})$ induced by pressure perturbations of Eq.~(\ref{e1})
can be found in a quite general form:
\begin{eqnarray}
\xi _{\mathbf{g}}(t)=-\frac{n_{s}g\tilde{V}_{g}}{\rho \cdot
\hat{\omega}_{g}} \int_{-\infty }^{t}\sin \left[
\hat{\omega}_{g}\left( t-t^{\prime }\right) \right] \times \nonumber \\
\times \exp \left[ -i\mathbf{g\cdot s}(t^{\prime })+\gamma
_{g}(t^{\prime }-t)\right] dt^{\prime },  \label{e4}
\end{eqnarray}
where $\hat{\omega}_{g}=\sqrt{\omega _{g}^{2}-\gamma _{g}^{2}}$.
Here we extend the treatment given in Ref.~\cite{MonKon-book} into
the range of a finite but small ripplon damping. For AC conditions
$\mathbf{s}(t)=\mathbf{s}_{0}\sin (\omega t)$. Introducing the new
variable $\tau =t^{\prime }-t$ and assuming $\omega \tau \ll 1$ we
represent $\mathbf{s}(t^{\prime })\approx
\mathbf{s}(t)+\mathbf{v}(t)\tau $. In this approximation surface
displacements can be found as
\begin{equation}
\xi _{\mathbf{g}}(t)=-\frac{n_{s}g\tilde{V}_{g}}{\rho \left[
 \omega _{g}^{2}-\left( \mathbf{gv}\right) ^{2}
-2i\gamma _{g}\mathbf{gv}\right] } e^{-i\mathbf{g\cdot s}(t)},
\label{e5}
\end{equation}
where $\mathbf{v=s}_{0}\omega \cos (\omega t)$ is the WS velocity.
This equation represents dimple sublattice moving in-phase with
the WS. In the limit $\mathbf{gv}\rightarrow 0$, Eq.~(\ref{e5})
surely gives the well known shape of steady dimples which is
independent of damping. If $\mathbf{gv}\sim \omega _{g}$, the
shape of dimples is affected by the WS velocity and damping. Thus,
under the condition $\omega \tau \ll 1$, which according to
Eq.~(\ref{e4}) requires $\omega \ll \gamma _{g}$, the dimple shape
changes with time continuously in such a way that it is always
adjusted to a given velocity $ \mathbf{v}(t)$. In other words, for
any fixed value of $\mathbf{v}(t)$ the dimple shape is the same as
that defined by the DC theory with the corresponding WS velocity.

For liquid $^{3}\mathrm{He}$, the weak ripplon-damping regime can
be realized only for superfluid phase at $T<0.3\,\mathrm{mK}$. In
this case, $ \gamma _{g}$ is determined by ballistic bulk
quasiparticle scattering from an uneven
interface~\cite{MonKon-05}:
\begin{equation}
\gamma _{q}=\frac{\hbar \left( k_{\mathrm{F}}\right) ^{4}}{8\pi
^{2}\rho } 2f(\Delta /T)q,  \label{e6}
\end{equation}%
where $k_{\mathrm{F}}$ is the Fermi momentum of quasiparticles in
liquid $ ^{3}\mathrm{He}$, $\Delta $ is the excitation gap, and
$f(x)=\left( e^{x}+1\right) ^{-1}$.  The ripplon damping of
superfluid $^{3}\mathrm{He}$ decreases with cooling at an
exponential rate. Still, in experiments on WS it can be\ just
reasonably small ($\gamma _{g}/\omega _{g}\sim 0.1$ or $0.01$).

In contrast, the ripplon damping in superfluid $^{4}\mathrm{He}$
is anomalously small. In the ballistic regime it is given
by~\cite{RocRogWil-96}
\begin{equation}
\gamma _{q}=\frac{\pi ^{2}}{60}\frac{\hbar }{\rho }\left(
\frac{T}{\hbar v_{ \mathrm{4He}}}\right) ^{4}q,  \label{e7}
\end{equation}
where $v_{\mathrm{4He}}$ is the first sound velocity. For
$n_{s}=10^{9}\,\mathrm{cm}^{-2}$ and $T=0.5\,\mathrm{K,}$ a simple
estimate gives $\gamma _{g_{1}}/\omega _{g_{1}}\sim 10^{-4}$.
Thus, for WS transport on superfluid $ ^{4}\mathrm{He}$ the
damping coefficient of ripplons is extremely small. It is
remarkable that this damping coefficient has the same dependence
on the wave-vector $q$ as that found for the ballistic regime of
liquid $^{3} \mathrm{He}$ [Eq.~(\ref{e6})].

In the reference frame moving along with the WS, the dimple
profile given by Eq.~(\ref{e5}) can be evaluated as
\begin{eqnarray}
\xi (\mathbf{r})=-\sum_{\mathbf{g}}\frac{n_{s}g\tilde{V}_{g}}{\rho
\left\vert D_{\mathbf{g}}(\mathbf{v})\right\vert ^{2}}\large{[}
\left( \omega _{g}^{2}-(\mathbf{gv})^{2}\right) \cos
(\mathbf{gr})- \nonumber \\
-2\mathbf{gv}\gamma _{g}\sin (\mathbf{gr})\large{]} ,\,\,\,\,\,
\label{e8}
\end{eqnarray}
where $D_{\mathbf{g}}(\mathbf{v})=\omega _{g}^{2}-\left(
\mathbf{gv}\right) ^{2}-i2 \mathbf{gv}\gamma _{g}$. At low
temperatures, especially for superfluid $^{3} \mathrm{He}$, the
summation over a large number of $\mathbf{g}$ is necessary to
ensure the convergence of the result. Consider $\mathbf{r}=(x,0)$
and assume that the driving force $-e\mathbf{E}$ is directed along
to the $x$-axis. Then, according to Eq.~(\ref{e8}), in the absence
of damping, dimples have a symmetrical shape with regard to
electron lattice sites. A finite damping introduces asymmetry in
the dimple shape due to the terms proportional to $\sin (g_{x}x)$.

\begin{figure}[tbp]
\begin{center}
\includegraphics[width=9.0cm]{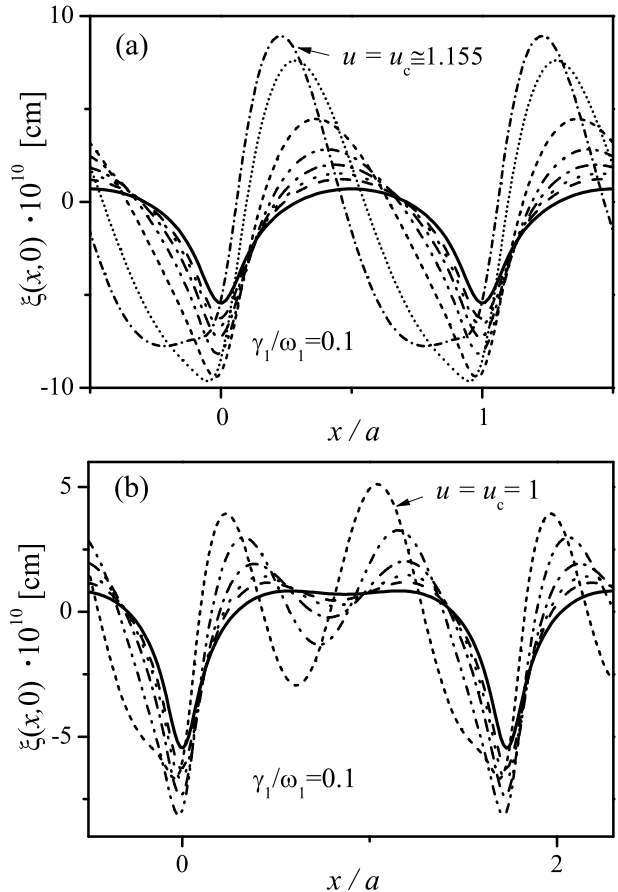}
\end{center}
\caption{Variations of the dimple sublattice profile $\xi (x,0)$
induced by the WS velocity for two typical velocity orientations:
NN-direction (a), and SN-direction (b). Steady dimples are shown
by the solid line. The dimensionless velocity increases from
$u=0.6$ to higher values by steps equal $0.1$. Calculations are
performed for superfluid $^{3}\mathrm{He}$,
$n_{s}=10^{8}\,\mathrm{cm^{-2}}$, $E_{\bot }=189\,\mathrm{V/cm}$,
and $T=0.277\,\mathrm{mK}$ ( $\gamma _{1}/\omega _{1}=0.1$).
 } \label{fig1}
\end{figure}

Shape variations of surface dimples induced by a finite WS
velocity are very sensitive to orientation of the vector
$\mathbf{v}$ with regard to symmetry axes of the WS. We shall
consider the following two typical directions. The direction of
the vector $\mathbf{v}$ which is parallel to the line connecting
two nearest neighbors of the electron lattice will be called
NN-direction. The direction of $\mathbf{v}$ which is parallel to
the line connecting second neighbors will be called SN-direction.
If the ripplon damping coefficient $\gamma _{1}$ is not too small
(about $0.1\cdot \omega _{1}$), the asymmetrical variations of the
dimple shape start already at substantial shifts from the B-C
resonance as shown in Fig.~\ref{fig1} ($a$ and $b$) for two
typical directions of the velocity vector. Here the dimple shape
of a motionless WS is shown by the solid line. The first
shape-line (dashed) of moving dimples is calculated for
dimensionless velocity $u=\mathrm{v}/\mathrm{v}_{1}=0.6$ (here
$\mathrm{v}_{1}=\omega _{g_{1}}/g_{1}$). For each next curve shown
in this figure, the parameter $u$ is increased by steps equal
$0.1$ except for the curve calculated at $u=2/\sqrt{\pi}$. The
last curve is calculated for the critical velocity $u_{c}$ of the
first B-C resonance, which equals $1$ for the SN-direction
(Fig.~\ref{fig1}b), and $2/\sqrt{\pi }$ for the NN-direction
(Fig.~\ref{fig1}a). For the symmetrical dimple shape, the average
force acting on electrons is obviously zero. The strong shape
asymmetry appears in order to transfer the kinetic friction acting
on the dimple sublattice by the environment to the electron
crystal.

\begin{figure}[tbp]
\begin{center}
\includegraphics[width=10.0cm]{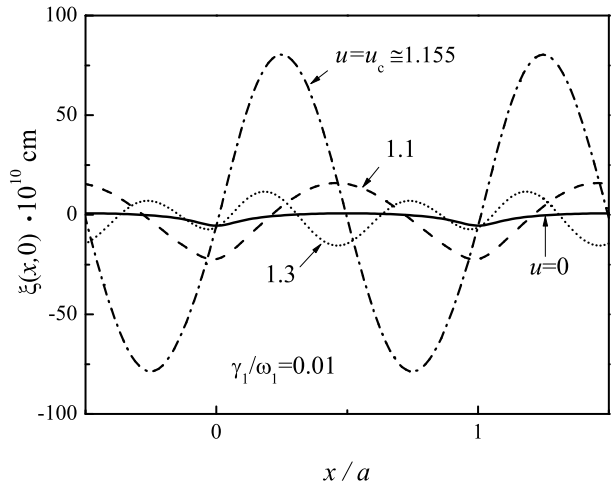}
\end{center}
\caption{Variations of the dimple sublattice profile $\xi (x,0)$
induced by the WS velocity near the first B-C resonance for very
small ripplon damping $\gamma _{1}/\omega _{1}=0.01$. The drift
velocity is oriented along the NN-direction. Calculations are
performed for superfluid $^{3}\mathrm{He}$,
$n_{s}=10^{8}\,\mathrm{cm^{-2}}$, $E_{\bot }=189\,\mathrm{V/cm}$,
and $T=0.2\,\mathrm{mK}$.} \label{fig2}
\end{figure}

For significantly smaller damping coefficient $\gamma
_{1}=0.01\cdot \omega _{1}$, which is expected for superfluid
$^{3}\mathrm{He}$ at $T=0.2\,\mathrm{mK}$, substantial shape
changes appear only if the WS velocity approaches the first B-C
resonance or exceeds it. Dimple shape variations which occur near
the vicinity of the first resonance are shown in Fig.~\ref{fig2}.
The important point is that the velocity-induced displacements
found for $ u=u_{c}=2/\sqrt{\pi }$ are much larger (about 15
times) than surface displacements in the initial surface dimple.
This means that dynamic decoupling of the WS from surface dimples
which shall be discussed below for a fixed field condition is
accompanied by creation of huge displacement waves moving in the
same direction.

At $u<u_{c}$, in spite of huge changes of the dimple profile the
average position of an electron remains the same being fixed to
the potential minimum formed by the dimple potential and the
driving electric field. For experimental conditions with a given
current, we may consider the evolution of surface dimple profile
even at $u>u_{c}$. For example, at $u=1.3$ surface displacements
induced by the WS velocity are already substantially reduced,
still their amplitude is larger than the initial dimple depth.

\begin{figure}[tbp]
\begin{center}
\includegraphics[width=10.0cm]{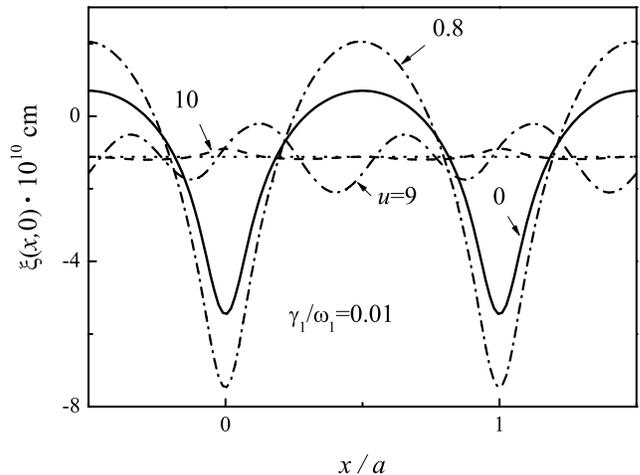}
\end{center}
\caption{Variations of the dimple sublattice profile $\xi (x,0)$
induced by the WS velocity away from the first B-C resonance
condition for $\gamma _{1}/\omega _{1}=0.01$. The drift velocity
is along the SN-direction. Other conditions are the same as in
Fig.~\ref{fig2}.} \label{fig3}
\end{figure}

It is instructive to consider dimple shape variations induced by
WS velocity which is away from the first B-C resonance condition.
The corresponding calculations are shown in Fig.~\ref{fig3}. As
expected, at $u=0.8<u_{c}$ the WS velocity just increases the
dimple depth. Significantly faster velocities $u\sim 10$ cause the
opposite effect: the surface displacements are substantially
reduced. For $u=9$ we accidentally reached one of higher B-C
resonances, causing strong asymmetry in the dimple shape. A small
detuning up to $u=10$ makes the shape nearly symmetrical which
means that the force transferred to the WS by dimples is close to
zero. It is instructive that in the limit $ u\rightarrow \infty $
surface displacements still remains as shown in Fig.~\ref{fig3} by
the straight dotted line, but instead of individual dimples we
have a row of interface valleys oriented in the direction of
motion.

Shape variations induced by B-C resonances lead to strong changes
in the associated mass of an electron dimple $M_{\mathrm{d}}$
which is given by
\begin{equation}
M_{\mathrm{d}}(\mathbf{v})=\frac{n_{s}}{\rho
}\sum_{\mathbf{g}}\frac{ g_{x}^{2}g\tilde{V}_{g}^{2}}{\left(
\omega _{g}^{2}-\left( \mathbf{gv} \right) ^{2}\right) ^{2}+\left(
2\gamma \mathbf{gv}\right) ^{2}},  \label{e9}
\end{equation}
where $\mathbf{v}$ is directed along the $x$-axis. For a finite
thickness of the liquid helium $d$, each term in the sum should be
multiplied by $\coth (gd)$. Obviously, the B-C resonances increase
the associated mass of a surface dimple. In the limiting case
$\mathbf{gv\rightarrow \infty }$, the associated mass
$M_{\mathrm{d}}(\mathbf{v})$ disappears because the dimple lattice
is rearranged in a raw of valleys (terms with $\mathbf{gv}=0$ do
not contribute in $M_{\mathrm{d}}$ because of the proportionality
factor $ g_{x}^{2}$).

\section{Field-velocity characteristics}

The asymmetry of the dimple shape with regard to the average
electron position in a lattice site causes a force $
\mathbf{F}^{(\mathrm{D})}$ acting on the electron solid. In
equilibrium this force is balanced by an external driving field.
By definition, $\mathbf{F}^{\mathrm{(D)}}$ is the sum of forces
acting on each electron $-\sum_{e}\partial V_{\mathrm{int}
}/\partial \mathbf{r}_{e}$ averaged over electron distribution
within the dimple [here $V_{\mathrm{int}}(\mathbf{r})$ is the
electron-ripplon interaction Hamiltonian whose Fourier transform
$V_{q}$ was used in Eq.~(\ref{e1})]. Electron distribution caused
by long wave-length fluctuations with $ \omega <\omega _{1}$
occurs together with surface dimples and therefore it should be
excluded from the averaging. Then, the average of the electron
density operator $n_{-\mathbf{q}}=\sum_{e}\exp \left(
i\mathbf{qr}_{e}\right) $ can be found as
\begin{equation}
\left\langle n_{-\mathbf{q}}\right\rangle _{\mathrm{f}}=N_{e}\exp
\left[ -q^{2}\left\langle u_{\mathrm{f}}^{2}\right\rangle
/4+i\mathbf{q}\cdot \mathbf{s}(t)\right] \delta
_{\mathbf{q},\mathbf{g}}.  \label{e10}
\end{equation}
Using this equation and Eq.~(\ref{e4}) in the general expression $
-\left\langle \sum_{e}\partial V_{\mathrm{int}}/\partial
\mathbf{r} _{e}\right\rangle _{\mathrm{f}}$, the force acting on
the WS can be written as
\begin{eqnarray}
F^{\mathrm{(D)}}(t)=-N_{e}\sum_{\mathbf{g}}\mathbf{g}\frac{n_{s}g\tilde{V}
_{g}^{2}}{\rho \hat{\omega}_{g}}\int_{0 }^{\infty }\sin \left(
\hat{\omega} _{g}\tau \right) e^{-\gamma _{g}\tau } \times \nonumber \\
\times \sin \left\{ \mathbf{g}\left[ \mathbf{s}
(t)-\mathbf{s}(t-\tau )\right] \right\} d\tau .  \label{e11}
\end{eqnarray}
For any spatially uniform displacement $\mathbf{s}(t)$ given, this
equation defines the in-plane force induced by the dimple
sublattice.

If the time scale of the WS displacement vector $\mathbf{s}(t)$ is
much longer than $\tau \sim 1/\gamma _{g}$, then $\mathbf{g}\left[
\mathbf{s}(t)-\mathbf{s}(t-\tau )\right] $ can be approximated by
$\mathbf{gv}(t)\tau $. In this limit $\mathbf{F}^{(\mathrm{D})}$
has the same form at that given by the DC treatment with the
constant velocity replaced by $\mathbf{v}(t)$. In equilibrium,
$F^{(\mathrm{D})}_{(x)}$ as well as the kinetic friction of the
electron lattice $F^{(\mathrm{fric})}_{(x)}=-N_{e}\nu
_{e}\mathrm{v}$ caused by electron scattering of other kinds are
balanced by the external force $ N_{e}eE$ (here we assume that
magnetic field is zero). The solution of the balance equation can
be represented as a field-velocity characteristic $E(
\mathrm{v})$:
\begin{equation}
E(\mathrm{v})=\frac{n_{s}}{e\rho
}\mathrm{v}\sum_{\mathbf{g}}\frac{g_{x}^{2}g
\tilde{V}_{g}^{2}2\gamma _{g}}{(\omega _{g}^{2}-g_{x}^{2}\mathrm{v}%
^{2})^{2}+4\gamma _{g}^{2}g_{x}^{2}\mathrm{v}^{2}}+\frac{m_{e}\nu _{e}}{e}%
\mathrm{v}.  \label{e12}
\end{equation}%
Using this equation, $E(\mathrm{v})$ can be calculated numerically
for any given damping coefficient and the collision frequency $\nu
_{e}$ caused by electron scattering with thermal ripplons, vapor
atoms, or even walls if WS is formed in a channel geometry.

For the DC case, Eq.~(\ref{e12}) is a two-dimensional extension of
the classical one-dimensional model of B-C scattering reported
previously~\cite{Vin-99} with the real ripplon damping parameter
and with more accurate electron-ripplon coupling. The driving
field found from the balance equation has sharp maxima in the
vicinity of B-C resonance conditions $g_{x}^{2}
\mathrm{v}^{2}\rightarrow \omega _{g}^{2}$. If the driving field
is given and the WS velocity is adjusted to the field, then
regions with $dE/d\mathrm{v}<0$ are unstable. This means that for
driving fields exceeding the major maximum of $E(\mathrm{v})$ the
balance of forces is not possible and the WS decouples from
surface dimples. According to Fig.~\ref{fig2}, decoupling of the
WS is accompanied by creation of huge surface waves moving with
the group velocity $u_{c}$.

In experiments on WS transport, usually it is the current which is
given, while the driving field is adjusted to the current by
electron redistribution which screens the external potential
variations. This is supported by the fact that regions with
$dE/d\mathrm{v}<0$ are experimentally
observed~\cite{ShiMonKon-04,GlaDotFoz-01}. Therefore,
field-velocity characteristics of electrons moving ultra-fast with
$u>u_{c}$ are very important for understanding the nonlinear WS
transport on superfluid helium.

For liquid $^{4}\mathrm{He}$ with $\gamma _{g}/$ $\omega _{g}\sim
10^{-4}$, Eq.~(\ref{e12}) applied to the DC case would give just a
set of extremely sharp peaks. At the same time, beyond these peaks
$E(\mathrm{v})$ is close to zero. For liquid $^{3}\mathrm{He}$,
the parameter $\gamma _{g}/$ $\omega _{g} $ can be much larger
than it is for liquid $^{4}\mathrm{He}$ (even larger than unity)
and B-C peaks of $E(\mathrm{v})$ can substantially overlap. It is
worth noting that $E(\mathrm{v})$ depends strongly on the velocity
vector orientation with regard to the 2D electron lattice. For
example, in the NN-direction of motion there is only one major
peak with $ \left\vert \mathbf{g}\right\vert =g_{1}$ at
$u=2/\sqrt{\pi }$ , while for the SN-direction there are two
equivalent peaks with $\left\vert \mathbf{g} \right\vert =g_{1}$
at $u=1$ and $u=2$.

In contrast to models of B-C scattering discussed previously, the
real experimental situation has one unavoidable complication: DC
measurements are practically impossible and by now all data are
obtained under AC conditions. This means that WS velocity and the
driving electric field are periodic functions of time with the
period $2\pi /\omega $. The frequency of experimental signal is
usually varies from $10^{4}\,\mathrm{s}^{-1}$ to about $ 6\cdot
10^{5}\,\mathrm{s}^{-1}$ which is much lower than the typical
ripplon frequency $\omega _{1}$. Therefore, it is conventionally
expected that the DC model of WS transport should give
qualitatively correct description of data obtained in such
experiments. We shall see that this is not true for two major
reasons.

First, we note that the condition $\omega \ll \omega _{1}$ is not
sufficient for adiabatic adjustment of surface displacements to WS
velocity variations. For example at the B-C resonance condition
($u=u_{c}$) we have a huge wave which follows the WS and for a
change to much smaller displacements a shape variation (without
changing the amplitude) is not sufficient. At the B-C resonance
the capillary wave accumulates great energy which should be
transferred to the environment in order to make the transition
from the surface displacements calculated for $ u=u_{c}\simeq
1.155$ (see Fig.) to the surface displacements calculated for $
u=1.3$ or even for $u=1.1$. This means that an adiabatic AC
extension of the DC model requires an additional restriction: the
frequency of the current should be much lower than the
corresponding ripplon damping ($\omega \ll \gamma _{1}$). This is
actually the condition which allowed us to transform
Eq.~(\ref{e11}) into Eq.~(\ref{e12}). For WS transport over
superfluid $^{3} \mathrm{He}$ this condition is satisfied even at
$\gamma _{g}/$ $\omega _{g}\sim 10^{-2}$. Regarding liquid
$^{4}\mathrm{He}$, the condition $\omega \ll \gamma _{1}$ requires
to use an AC frequency which is much lower than $
10^{4}\,\mathrm{s}^{-1}$.

Secondly, even if the above noted condition is fulfilled,
Eq.~(\ref{e12}) cannot be used directly for plotting field
velocity characteristics. Under AC conditions, it is important
which quantities are actually measured and presented in the
field-velocity characteristics. If time averaged quantities are
considered, in the nonlinear regime the outcome can be
qualitatively different for different kinds of averaging. For
example, even for harmonic velocity $u_{0}\cos \omega t$, the
mean-square time averaging of the driving field
$\sqrt{\left\langle E^{2}\right\rangle }$ and averaging of the
absolute value $\left\langle \left\vert E\right\vert \right\rangle
$ give qualitatively different results for the function
$E(u_{0}\cos \omega t)$ defined by Eq.~(\ref{e12}). This could be
easily proven by considering the right-side tail of the B-C
resonance in the limiting case $\gamma _{g}\rightarrow 0$. For
averaging $\left\vert E\right\vert $, in this limit a part of the
corresponding integrand can be rearranged as the $\delta $
-function [see Eq.~(\ref{e12})], and the final result will not
depend on $ \gamma _{g}$. In contrast, the integrand of the
quantity $\left\langle E^{2}\right\rangle $ is squared, and,
therefore, the resonance tail of $ \sqrt{\left\langle
E^{2}\right\rangle }$ increases with reducing $\gamma _{g} $.

As a measure of the alternating field one can choose the main term
of the Fourier series representing $E(t)$:
\begin{equation}
E_{\omega }=\frac{\omega }{\pi }\int_{-\pi /\omega }^{\pi /\omega
}E(t)\cos \left( \omega t\right) dt.  \label{e13}
\end{equation}%
It is also a kind of time averaging, and similar to $\left\langle
\left\vert E\right\vert \right\rangle $ it has a finite resonance
tail in the limiting case $\gamma _{g}\rightarrow 0$. In the
following we shall consider only quantities $E_{\omega }$ and
$\left\langle \left\vert E\right\vert \right\rangle $ which will
be used for presenting the field-velocity relationship.

Assume that the conditions of given current is realized and
$u(t)=u_{0}\cos \omega t$. Using the adiabatic treatment discussed
above, we insert $\mathrm{v}(t)\mathrm{=v}_{1}u(t)$ into
Eq.~(\ref{e12}) and evaluate the time integral of Eq.~(\ref{e13}).
It is convenient to introduce two integer variables $m$ and $n$ to
describe the reciprocal lattice vectors
$\mathbf{g}_{m,n}=m\mathbf{g}^{(1)}+n\mathbf{g}^{(2)}$ (here
$\mathbf{g}^{(1)}$ and $\mathbf{g}^{(2)}$ are primitive vectors of
this lattice). Then, after some algebra, the Fourier transform
$E_{\omega }$ can be represented as a function of the velocity
amplitude
\begin{eqnarray}
E_{\omega }(u_{0})=\mathrm{v}_{1}\frac{n_{s}\rho ^{1/2}}{e\alpha
^{3/2}g_{1}}
\sum_{m,n}\frac{p_{m,n}\tilde{V}_{g_{m,n}}^{2}}{g_{m,n}^{1/2}}
Q(p_{m,n}u_{0},\beta _{m,n})+ \nonumber \\
+\frac{m_{e}\nu _{e}}{e}\mathrm{v}_{1}u_{0}. \,\,\,\, \,\,\,\,
\label{e14}
\end{eqnarray}
Here we use the following dimensionless notations
\begin{equation}
p_{m,n}=\frac{\sqrt{\left( \mathbf{g}_{m,n}\right)
_{x}^{2}}}{g_{1}}\left( \frac{g_{1}^{3}}{g_{m,n}^{3}}\right)
^{1/2},\,\,\,\,
\beta _{m,n}=2 \frac{\gamma _{g_{m,n}}}{\omega _{g_{m,n}}},
\label{e15}
\end{equation}%
\begin{equation}
Q(w,\beta )=\frac{4w\beta }{\pi }\int_{0}^{\infty
}\frac{dy}{\left[ \left( 1+ y^{2}-w^{2}\right) ^{2}+w^{2}\beta
^{2}\left( 1+y ^{2}\right) \right] }.  \label{e16}
\end{equation}%
The integral $Q(w,\beta )$ can be evaluated in an analytical form:%
\begin{equation}
Q(w,\beta )=\frac{2\beta}{wG}\mathrm{Re}\left[
\frac{1}{\sqrt{w^{2}-1-w^{2}\beta ^{2}/2-\mathrm{i}w^{2}G}}\right]
,  \label{e17}
\end{equation}
where $G\left( \beta \right) =\beta \sqrt{ 1-\beta ^{2}/4 }$.
Remarkably, in the limit $\gamma _{g}\rightarrow 0$ ($\beta
\rightarrow 0$) for $ w>1$ there is a finite asymptote
\begin{equation}
Q(w,\beta )\rightarrow \frac{2\theta (w-1)}{w\sqrt{w^{2}-1}},
\label{e18}
\end{equation}%
where $\theta (x)$ is the unit step-function.

For $\left\langle \left\vert E\right\vert \right\rangle $ as a
function of the dimensionless velocity amplitude $u_{0}$, an
equation similar to Eq.~(\ref{e14}) is found. The only difference
which appears for such averaging is that instead of $Q(w,\beta )$
we should use another function $I(w,\beta )$
defined by%
\begin{equation}
I(w,\beta )=\frac{2\beta }{\pi G}\mathrm{Re}\left[ \frac{\arctan \left[ \frac{%
\mathrm{i}w}{\sqrt{ w^{2}-1+\beta ^{2}/2 +\mathrm{i}G}}\right]
}{\sqrt{ w^{2}-1+\beta ^{2}/2 +\mathrm{i}G}}\right], \label{e19}
\end{equation}%
As expected in the limiting case $\gamma _{g}\rightarrow 0$
($\beta \rightarrow 0 $), it also has a finite asymptote
\begin{equation}
I(w,\beta )\rightarrow \mathtt{\ }\frac{\theta
(w-1)}{\sqrt{w^{2}-1}}, \label{e20}
\end{equation}%
which means that the right-side tails of B-C resonances are independent of
small damping.

\begin{figure}[tbp]
\begin{center}
\includegraphics[width=10.0cm]{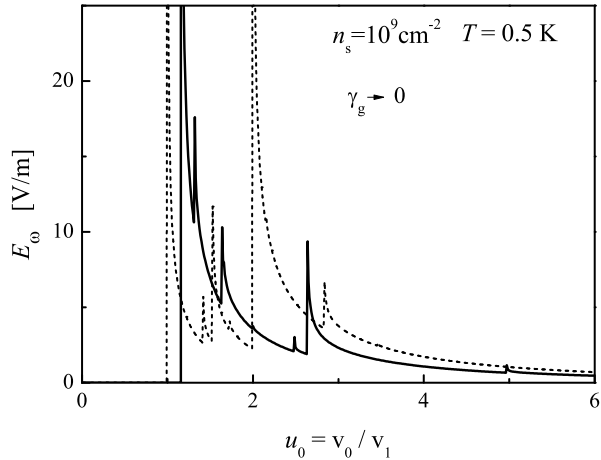}
\end{center}
\caption{The first Fourier coefficient of the driving field
$E_{\omega }$ vs the drift velocity amplitude for two typical
orientations of the velocity: NN-direction (solid line) and
SN-direction (dashed line). Calculations where performed for
superfluid $^{4}\mathrm{He}$, $n_{s}=10^{9}\,\mathrm{cm^{-2}}$,
$T=0.5\,\mathrm{K}$, assuming the ripplon damping is set to zero.
} \label{fig4}
\end{figure}

Already from the analysis of the integrals $Q(w,\beta )$ and
$I(w,\beta )$ given above one can conclude that B-C resonance
tails of the AC treatment differ (even qualitatively) from that
found for the DC models. The left-side tail becomes even steeper
for both $E_{\omega }(u_{0})$ and $\left\langle \left\vert
E\right\vert \right\rangle $, while the right-side tails extend
far beyond the resonance and are independent of ripplon damping in
the limit $\gamma _{g}\rightarrow 0$. This behavior is illustrated
in Fig.~\ref{fig4} where $ E_{\omega }(u_{0})$ is plotted for two
typical directions of the WS velocity [NN-direction (solid line),
and SN-direction (dashed line)] assuming $\gamma _{g}\rightarrow
0$. The other parameters are taken for the liquid
$^{4}\mathrm{He}$ case. Thus, instead of $\delta $-peaks of the
classical B-C scattering model here we have saw-tooth shaped peaks
with long right-side tails. As noted above, field-velocity
characteristics depend strongly on the direction of the WS
velocity. It is interesting that for the SN-direction there are
two major B-C peaks and the second one (at $u_{0}\rightarrow 2$)
becomes even more prominent than the first one because in the AC
cause at $ u_{0}>2$ the velocity sweeps through the both
resonances. Of course, considering the limiting case $\gamma _{g}
/\omega _{1}\rightarrow 0$ we should keep in mind that $\omega $
should be much smaller than $\gamma _{g}$. Therefore, real ripplon
damping should be taken into account for consistent analysis of
field-velocity relationships induced by B-C scattering.

For nonlinear WS transport over superfluid $^{4}\mathrm{He}$, the
ripplon damping parameter is extremely small and the condition
$\gamma _{g}< \omega \ll \omega _{g}$ is realized in most of known
experiments. This case requires a special treatment because large
surface displacements excited at a first instance of $u(t)=u_{c}$
cannot be relaxed back within the period of current oscillations.
Therefore a new steady regime will be developed for each AC
frequency, which can be far away from the solutions found above
for the condition $\omega \ll \gamma _{g}$.

We have to return back to the exact solution of Eq.~(\ref{e11}).
In this equation now we insert $s_{x}(t)=(\mathrm{v}_{0}/\omega )
\sin \left( \omega t\right) $ and then perform the time averaging
defined by Eq.~(\ref{e13}). Here we disregard the unimportant
correction induced by $\nu _{e}$. Then quite generally, $E_{\omega
}(\mathrm{v}_{0})$ can be found as
\begin{eqnarray}
E_{\omega
}(\mathrm{v}_{0})=\sum_{\mathbf{g}}\frac{g_{x}n_{s}g\tilde{V}
_{g}^{2}}{e\rho \hat{\omega}_{g}}2\int_{0}^{\infty }\sin \left(
\hat{\omega} _{g}\tau \right) e^{-\gamma _{g}\tau }\cos \left(
\frac{\omega \tau }{2}
\right)  \nonumber  \\
\times J_{1}\left[ 2g_{x}\frac{\mathrm{v}_{0}}{\omega }\sin \left(
\frac{ \omega \tau }{2}\right) \right] d\tau , \hspace{2cm}
\label{e21}
\end{eqnarray}%
where  $J_{1}(z)$ is the Bessel function of the first kind. For
$E_{\omega }$ as a function of the dimensionless velocity
amplitude $u_{0}$, an equation similar to Eq.~(\ref{e14}) can be
found. In the general case, the integral $Q(w,\beta )$ of
Eq.~(\ref{e14}) should be replaced by $Q(w,\beta ,\omega ^{\prime
})$
defined as%
\begin{eqnarray}
Q(w,\beta ,\omega ^{\prime })=4\int_{0}^{\infty }\sin \left(
2y\right) e^{-\beta y}\cos \left( \omega ^{\prime }y\right)
\times \nonumber \\
\times J_{1}\left[ 2\frac{w}{\omega ^{\prime }}\sin \left( \omega
^{\prime }y\right) \right] dy, \hspace{2cm} \label{e22}
\end{eqnarray}%
where $\omega ^{\prime }=\omega /\hat{\omega} _{g}$. The
dimensionless function $Q(w,\beta ,\omega ^{\prime })$ describes
the shape of a single B-C resonance of $E_{\omega }(u_{0})$ for
arbitrary current frequency and ripplon damping (here $w\propto
u_{0}$ and $\beta \propto \gamma _{g}$). It is easy to see that in
the limiting case $\omega ^{\prime }\rightarrow 0$ and $\beta
\rightarrow 0$ analyzed above, Eq.~(\ref{e22}) provides the
correct asymptote shown in Eq.~(\ref{e18}). For conditions $\gamma
_{g}< \omega \ll \omega _{g},$ and $ g_{x}\mathrm{v}_{0}>1$, the
argument of the Bessel function entering the integrand of
Eq.~(\ref{e21}) attains huge numbers because $\sin \left( \omega
\tau /2\right) \sim 1$. This can lead to remarkable field-velocity
relationships with side-oscillations which we shall discuss in the
following.

\section{Results and discussions}

\begin{figure}[tbp]
\begin{center}
\includegraphics[width=10.0cm]{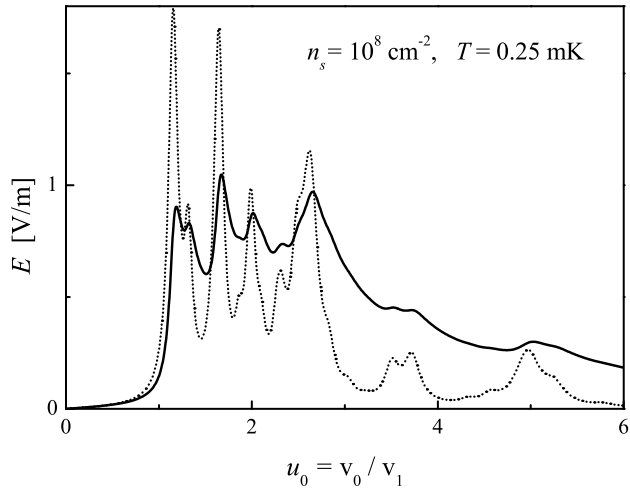}
\end{center}
\caption{Field velocity relationship for the DC case (dashed
line), and the first Fourier coefficient of the driving field
$E_{\omega }$ vs the drift velocity amplitude for the AC case
(solid line). Drift velocity is oriented along the NN-direction.
Calculations where performed for superfluid $^{3}\mathrm{He}$,
$n_{s}=10^{8}\,\mathrm{cm^{-2}}$, $T=0.25\,\mathrm{mK}$. }
\label{fig5}
\end{figure}

Consider briefly the velocity-field relationship $E_{\omega }(u)$
for WS transport over superfluid $^{3}\mathrm{He}$ at
$T=0.25\,\mathrm{mK}$. At such a temperature the damping
coefficient $\gamma _{g_{1}}\simeq 1.54\cdot 10^{6}\,
\mathrm{s}^{-1}$ is much lower than the B-C resonance frequency
$\omega _{1}$ and is still much higher than the typical current
frequency $\omega $ used in experiments. The later condition makes
the adiabatic AC extension of the B-C scattering model introduced
here applicable. The results of numerical evaluations of
Eq.~(\ref{e14}) are shown in Fig.~\ref{fig5}. The velocity-field
characteristic given by the DC classical model is shown by the
dotted line. It consists of series of B-C peaks with nearly
symmetric tails. For the first Fourier coefficient as a function
of the drift velocity amplitude, the AC theory gives lower peaks
which are asymmetric with regard to the maximum positions. It is
important that time averaging used in evaluation of $E_{\omega
}(u_{0})$ do not smooth out the B-C resonances completely. Another
important feature of the AC treatment discussed here is the
appearance of long right-side tails of the B-C resonances. The
left-side tails ($u_{0}<u_{c}$) become even more steeper, because
under AC conditions electrons spend only a little time near the
B-C resonance.

\begin{figure}[tbp]
\begin{center}
\includegraphics[width=10.0cm]{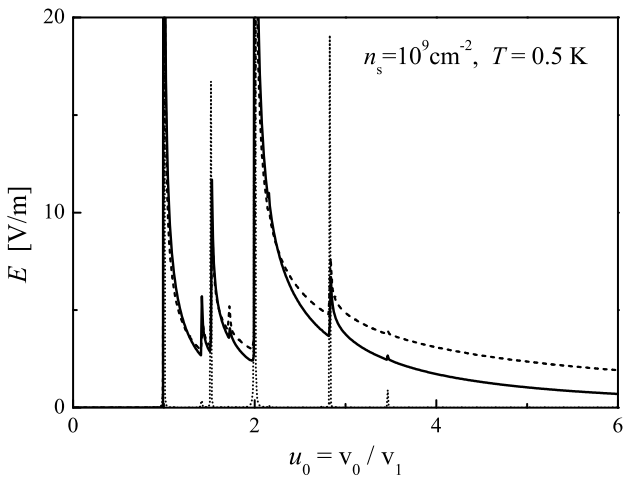}
\end{center}
\caption{Field velocity relationship for two kinds of averaging of
the alternative driving field: $E_{\omega}$ (solid line) and $\pi
\left\langle \left\vert E\right\vert \right\rangle /2$ (dashed
line). Drift velocity is oriented along the SN-direction.
Calculations where performed for superfluid $^{4}\mathrm{He}$,
$n_{s}=10^{9}\,\mathrm{cm^{-2}}$, and $T=0.5\,\mathrm{K}$ with the
ripplon damping parameter defined by Eq.~(\ref{e7}). }
\label{fig6}
\end{figure}

The most interesting experimental results on nonlinear WS
transport were obtained employing liquid
$^{4}\mathrm{He}$~\cite{ShiKon-95,KriDjeFoz-96}. In this case even
for $T\sim 0.5\,\mathrm{K}$ the ripplon damping parameter given by
Eq.~(\ref{e7}) is extremely small. First, we assume that $\omega $
is low enough to make Eq.~(\ref{e14}) applicable. The comparison
of results obtained for the two kinds of averaging of the electric
field ($E_{\omega }$ and $\left\langle \left\vert E\right\vert
\right\rangle $) is given in Fig.~\ref{fig6}, assuming that the
velocity is along the NN-direction and $\nu _{e}=0$. Here the
sharp peaks of the DC model (dotted line) are strongly smoothed by
the time averaging of the AC model ($E_{\omega }$ is shown by the
solid line, and $ \left\langle \left\vert E\right\vert
\right\rangle $ - by the dashed line). Additionally, the maximum
values of the B-C peaks are greatly reduced as compared to the
results calculated for the DC case (dotted line).

\begin{figure}[tbp]
\begin{center}
\includegraphics[width=10.0cm]{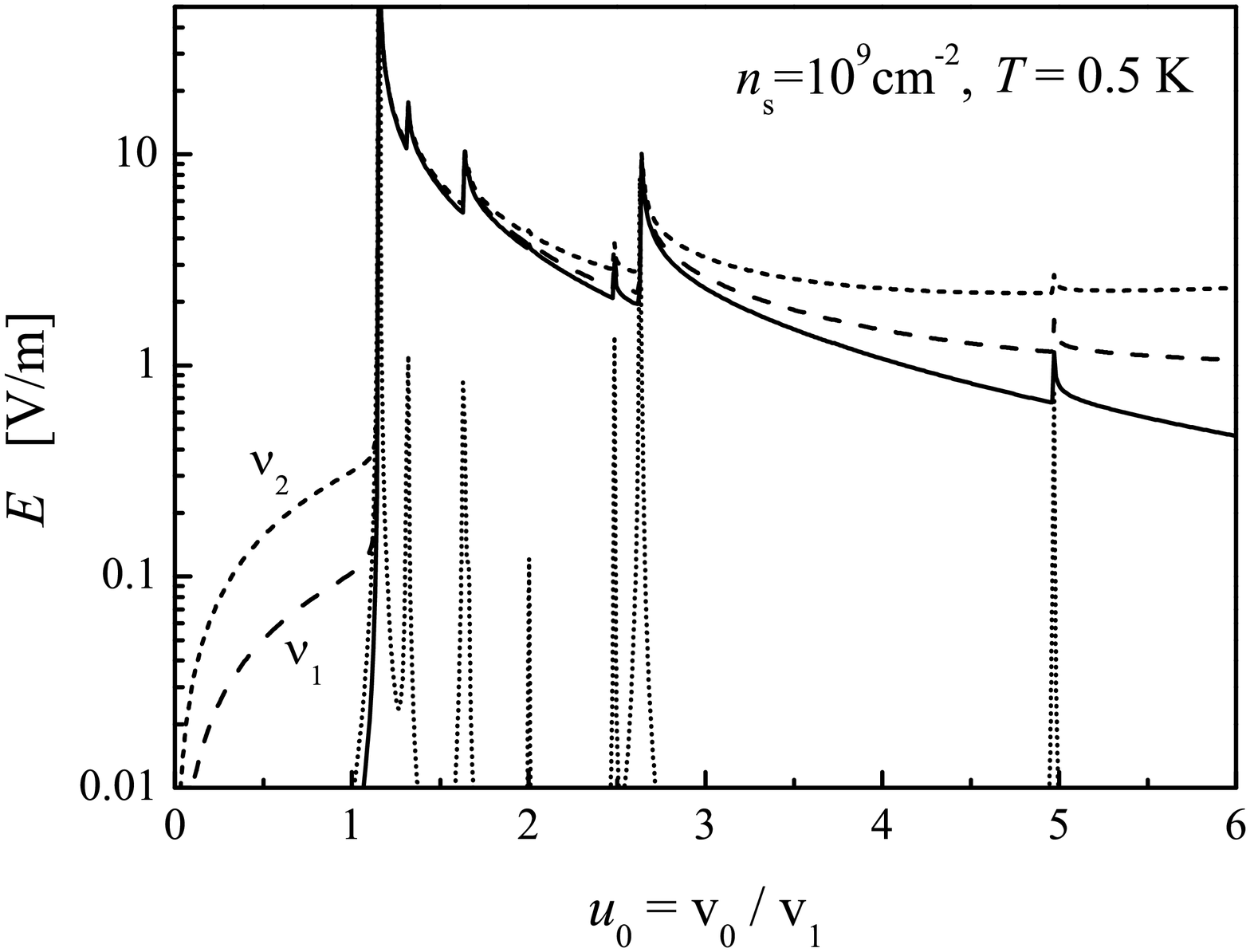}
\end{center}
\caption{The main Fourier coefficient of the driving field
$E_{\omega }$ vs the drift velocity amplitude for different values
of the electron collision frequency $\nu _{e}$: $0$ (solid line),
$2.4\cdot 10^{9}\,\mathrm{s}^{-1}$ (dashed line), and $7.5\cdot
10^{9}\,\mathrm{s}^{-1}$ (short dashed line). Drift velocity is
oriented along the NN-direction. The DC case results are shown by
the dotted line. Calculations where performed for superfluid
$^{4}\mathrm{He}$, $n_{s}=10^{9}\,\mathrm{cm^{-2}}$,
$T=0.5\,\mathrm{K}$. } \label{fig7}
\end{figure}

The influence of a finite electron collision frequency $\nu _{e}$
due to scattering with thermal ripplons and walls is analyzed for
the SN-direction and shown in Fig.~\ref{fig7}. A reasonable
estimate for the electron collision frequency $\nu _{e}\simeq \nu
_{1}=2.4\cdot 10^{9}\,\mathrm{s}^{-1}$ (for chosen $n_{s}$,
$E_{\bot }$ and $T$) is found considering electron scattering with
thermal ripplons in the usual way~\cite{MonKon-book} and taking
into account that at low temperatures the average kinetic energy
of an electron in the WS state differs substantially from $T$. For
experiments with WS in the channel geometry, $\nu _{e}$ can be
even higher because of the WS friction at the channel walls. In
order to illustrate this effect is Fig.~\ref{fig7} we considered
also a larger value $\nu _{e}\simeq \nu _{2}=7.5\cdot
10^{9}\,\mathrm{s}^{-1}$. According to this figure, the electron
collision frequency $\nu _{e}$ affects the both tails of the field
velocity characteristic. At the left side $\nu _{e}$ and time
averaging of Eq.~(\ref{e13}) act in the opposite ways. The
left-side tail ($u_{0}<u_{c}$) becomes less steep for a finite
$\nu _{e}$. At the right side $\nu _{e}$ acts in the same way as
the time averaging, increasing the right-side tail and making $
E_{\omega }(u_{0})$ more flatter in the region $u_{0}>u_{c}$. In
general, due to the both these effects the field-velocity
characteristic of the WS acquires a distinctive "nose" shape.

\begin{figure}[tbp]
\begin{center}
\includegraphics[width=10cm]{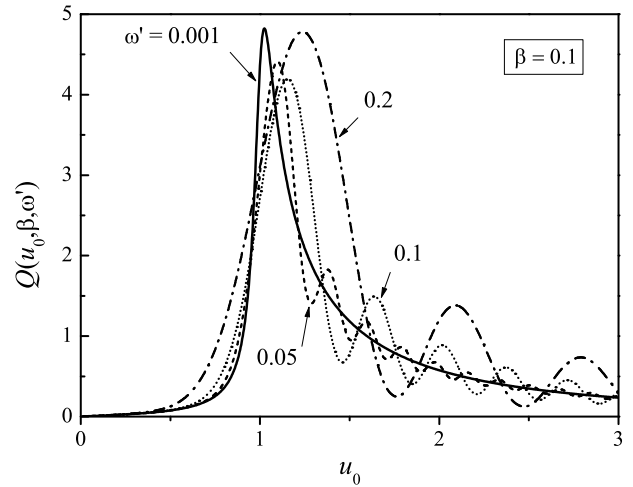}
\end{center}
\caption{Shape transformations of the function $Q(u_{0},\beta
,\omega ^{\prime})$ which describes a single B-C resonance under
AC conditions. Calculations are performed for current frequencies:
$\omega ^{\prime}= 0.001$ (solid line), $0.05$ (dashed line),
$0.1$ (dotted line), $0.2$ (dash-dotted line), $0.3$
(dash-dot-dotted line).} \label{fig8}
\end{figure}

The numerical calculations presented in Figs.~\ref{fig6} and
\ref{fig7} were done assuming $\omega \ll \gamma _{g_{1}}$. In
experiments on WS transport over superfluid $^{4}\mathrm{He}$ this
condition was not realized. The influence of the condition $\omega
>\gamma _{g_{1}}$ on the field-velocity characteristics can be
understood using the general expressions for $E_{\omega
}(\mathrm{v}_{0})$ and $Q(w,\beta ,\omega ^{\prime })$ given in
Eqs.~(\ref{e21}) and (\ref{e22}). The main features of B-C
scattering under AC conditions can be revealed from the
dimensionless function $Q(w,\beta ,\omega ^{\prime })$ which
describes the shape of a single B-C resonance [see
Eq.~(\ref{e14})]. Consider the main B-C resonance when we can set
$w=u_{0}$, $\beta =2\gamma _{1}/\omega _{1}$ and $\omega ^{\prime
} =\omega /\omega _{1}$, and for simplicity assume $\beta =0.1$.
In the limiting case $\omega ^{\prime } \ll \beta $, the function
$Q(u_{0},\beta ,0)$ coincides with $Q(u_{0},\beta )$ obtained in
Eq.~(\ref{e17}). For example, it is practically impossible to
distinguish $Q(u_{0},\beta ,0.001)$ shown in Fig.~\ref{fig8} by
the solid line from $Q(u_{0},\beta )$ given by Eq.~(\ref{e17}). As
a function of the dimensionless velocity, $Q(u_{0},\beta ,0.001)$
has a typical saw-tooth shape discussed above.

Remarkable shape transformations of $Q(u_{0},\beta ,\omega
^{\prime })$ as a function of $u_{0}$ occur when $\omega ^{\prime
}$ approaches and exceeds the value of the parameter $\beta $
which is proportional to ripplon damping. A sharp (from the
left-side) saw-tooth peak of a single B-C resonance is developed
into distinctive smooth oscillations which [according to
Eq.~(\ref{e14})] result in similar oscillations of
$E_{\omega}(u_{0})$. The amplitude and the period of oscillations
are gradually increase with $\omega ^{\prime }$ in the range
considered. These oscillations represent a new regime of B-C
scattering of the WS which occur under the AC condition, when the
current frequency becomes comparable or larger than the ripplon
damping.

\begin{figure}[tbp]
\begin{center}
\includegraphics[width=10cm]{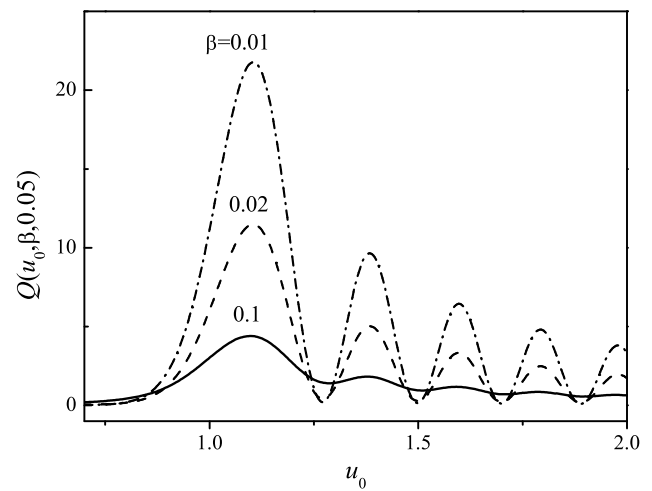}
\end{center}
\caption{Shape transformations of the function $Q(u_{0},\beta
,\omega ^{\prime})$ which describes a single B-C resonance under
AC conditions for a fixed frequency ($\omega ^{\prime }=0.05$) and
different ripplon-damping coefficients: $\beta= 0.1$ (solid line),
$0.02$ (dashed line), and $0.01$ (dash-dotted line).} \label{fig9}
\end{figure}

The period of new conductivity oscillations depends on the
relation between the frequency of the current $\omega$ and the
frequency of ripplons excited $\omega _{g}$. According to
Fig.~\ref{fig8}, it increases with the ratio $\omega /\omega
_{g}$. The ripplon damping just increases the amplitude of
oscillations. The later is illustrated in Fig.~\ref{fig9} where
$Q(u_{0},\beta ,\omega ^{\prime })$ is plotted vs $u_{0}$ for a
fixed value of the ratio $\omega /\omega _{1}=0.05$ and different
values of the damping parameter.

There are other important points which follow from
Figs.~\ref{fig8} and \ref{fig9}. For low damping the first maximum
of the field-velocity characteristic can be substantially larger
than the B-C peak value found in the limiting case $\omega \ll
\gamma _{1}$. Secondly, due to the finite frequency $\omega $,
even for a very small ripplon damping the left-side tail is not
steep as it was for $\omega \ll \gamma _{1}$. Additionally, the
maximum position is substantially shifted to higher drift
velocities.

The physics of side-oscillations in the field-velocity
relationship can be explained as follows. If $u_{0}>u_{c}$, then
even during a period of current oscillations the WS passes through
the B-C scattering point four times. The phase difference between
surface waves of the same $\mathbf{q}=\mathbf{g}$ excited at
different times increases with the velocity amplitude. Thus,
depending on the velocity amplitude $u_{0}$ the excited waves can
interfere constructively or destructively, which is the reason for
the side-oscillations. The higher frequency of the current, the
larger amplitude is necessary to produce the same
phase-difference.

Most of experiments on the nonlinear WS transport are performed
using a Corbino geometry of which the alternating current is
spatially nonuniform, and only a limited area of the WS can
satisfy the B-C scattering conditions. This area changes with time
because of the AC conditions. This experimental situation is very
difficult to analyze. One may conclude that spatial variations of
the current would additionally smooth out the B-C resonances. It
should be noted that the field-velocity characteristics of the WS
with a "nose" shape without the B-C peaks where observed in the
experiment on WS transport in a channel
geometry~\cite{GlaDotFoz-01}. It is interesting that conductance
oscillations similar to the B-C oscillations shown in
Fig.~\ref{fig8} were also reported in this experiment. Therefore,
our theoretical results give an alternative explanation for
oscillations in electronic response observed for SEs on superfluid
helium $^{4}\mathrm{He}$.

\section{Conclusions}

In summary, we have analyzed the nonlinear WS transport over
superfluid $^{3}\mathrm{He}$ and $^{4}\mathrm{He}$ under AC
conditions. The theory developed for a spatially uniform
alternating current indicates that the field-velocity relationship
obtained previously in the classical DC model of B-C scattering is
not applicable for time averaged quantities such as the first
Fourier coefficient. The detailed analysis is given for two
important limiting cases of low and high frequency of the electron
current. For frequencies which are much lower than the ripplon
damping coefficient, calculations based on the new theory lead to
asymmetric B-C peaks of a saw-tooth shape which are strongly
broadened at the right side. The broadening of the right-side
tails do not depend on small ripplon damping. The left-side tails
of the B-C resonances become even steeper which preserves the main
B-C anomaly in the field-velocity characteristic.

For current frequencies which are comparable with the ripplon
damping or even higher, the new nonlinear regime of B-C scattering
of the WS is reported. In this regime each B-C peak is transformed
into an oscillatory field-velocity relationship due to
interference of ripplons multiply excited at different times. The
evolution of surface displacements of the dimple sublattice with
increasing the current amplitude calculated in this work, as well
as the new field-velocity relationships obtained for alternating
current, help to understand the nonlinear conductivity of the WS
on superfluid helium observed in different experiments.

\section{Acknowledgments} The work is partly supported by the Grant-in-Aids for
Scientific Research from Monka-sho.

\end{document}